# Characterizing and Comparing COVID-19 Misinformation Across Languages, Countries and Platforms

Golshan Madraki[1], Isabella Grasso, Jacqueline M. Otala, Yu Liu, Jeanna Matthews

## Abstract

Misinformation/disinformation about COVID-19 has been rampant on social media around the world. In this study, we investigate COVID-19 misinformation/ disinformation on social media in multiple languages/countries: Chinese (Mandarin)/China, English/USA, and Farsi (Persian)/Iran; and on multiple platforms such as Twitter, Facebook, Instagram, WhatsApp, Weibo, WeChat and TikTok. Misinformation, especially about a global pandemic, is a global problem yet it is common for studies of COVID-19 misinformation on social media to focus on a single language, like English, a single country, like the USA, or a single platform, like Twitter. We utilized opportunistic sampling to compile 200 specific items of viral and yet debunked misinformation across these languages, countries and platforms emerged between January 1 and August 31. We then categorized this collection based both on the topics of the misinformation and the underlying roots of that misinformation. Our multi-cultural and multi-linguistic team observed that the nature of COVID-19 misinformation on social media varied in substantial ways across different languages/countries depending on the cultures, beliefs/religions, popularity of social media, types of platforms, freedom of speech and the power of people versus governments. We observe that politics is at the root of most of the collected misinformation across all three languages in this dataset. We further observe the different impact of government restrictions on platforms and platform restrictions on content in China, Iran, and the USA and their impact on a key question of our age: how do we control misinformation without silencing the voices we need to hold governments accountable?

## 1. Introduction

Along with the COVID-19 pandemic crisis, an infodemic (World Health Organization, 2020) crisis has affected all aspects of human lives from elections to public health response around the world. Social media has played a critical role in this infodemic crisis. First, social media offers a free and easy-access platform for users to share content (both true and false) in the form of posts, videos, pictures, and memes, all with a wide range of audiences (Weinberger, 2011). Second, the COVID-19 outbreak has forced people around the world to be quarantined and consequently interactions have shifted away from face-to-face interactions and even more towards online/social media interactions. This means that more people are exposed to unreliable content circulating on social media and many users struggle with distinguishing between facts and lies/fictions about COVID-19.

Since the beginning of the pandemic, some official efforts have been implemented to debunk these lies and inaccurate information circulating on social media, despite substantial disagreement about which corrective measures for fact checking are practical and appropriate for massive social media platforms. Still, these efforts seem not enough and there is a widespread consensus that an integrated sustainable global effort is warranted across different languages and through different platforms as are targeted in our study (Dizikes, 2020; Pennycook et al., 2020a).

In the literature, the problematic information has been categorized into misinformation, disinformation, and mal-information, with some debate about definitions (Karduni, 2019; Wardle, 2017). In this paper, we consider the following definitions:

**Misinformation** is incorrect information created without the intention of causing harm (e.g. encouraging people to wear a face shield without realizing how ineffective it could be).
**Disinformation** is incorrect information and intentionally created to hurt an individual, a group, or a country (e.g. a drug company spreading out a false rumor that its new medicine can cure COVID-19 with an intent to increase profit.)
**Mal-information** is correct information (based on reality), but used to cause harm to an individual, a group, or a country (e.g. justifying the high rate of confirmed cases by claiming that it is because of increasing the rate of testing.) (Infodemic Toolkit, 2020; Kujawski, 2019; Wardle, 2019).

Distinguishing between these three categories, particularly misinformation and disinformation, can be difficult or even impossible in some cases as can require assessing the intent of the creator. Thus, more broadly, here in this paper, we will use the general term 'misinformation' to refer to all three categories. According to a study from Oxford University, misinformation is also often true information reconfigured or recontextualize and less commonly fabricated (Brennen et al., 2020). This reconfigured information can circulate even faster than fully fabricated

---

[1] Corresponding Authors: Clarkson University. Email: gmadraki@clarkson.edu.

stories, taking in 87% of the interactions in one study (Brennen et al., 2020).

In this paper, COVID-19 misinformation, broadly defined to include disinformation and mal-information as well, will be investigated within multiple languages (Chinese, English, and Farsi) about multiple countries (China, Iran, and the USA) on different social media platforms. The COVID-19 pandemic broke out first in Wuhan, China in December 2019. Then, Iran became a hotspot in February 2020. The USA has been a clear hotspot as well with 22% of confirmed world-wide COVID cases despite only 4.25% of the world population (Dong et al., 2020). We chose these three countries as a lens through which to consider major differences between COVID-19 misinformation around the world. We show how examining social media misinformation from the perspective of one country, one language, or one platform, misses important and more holistic aspects of the pandemic. Through this research, we build a more comprehensive picture of how misinformation has exacerbated the COVID-19 crisis around the world.

Our research team includes native speakers of Chinese, English, and Farsi who were born and raised in China, Iran, and the USA. To the best of our knowledge, we are the first to examine the multilingual social media landscape by using the opportunistic sampling method to collect a dataset of verified and viral COVID-19 misinformation across 3 languages: Chinese, English, and Farsi. Our multicultural and multilingual team observed that the nature of COVID-19 misinformation on social media varies in substantial ways across different languages/countries depending on the cultures, beliefs, religions, popularity of social media, types of platforms, freedom of speech, the power of people versus governments, etc. Based on these observations, we proposed a novel and comprehensive categorization of the COVID-19 misinformation based on their topics and roots such that these categories are all relevant and extendable in all three languages.

It is worth noting the difference in government policies for controlling misinformation in Iran, China, and the USA. China has strict government control over which platforms can be used and enforces these controls. Approved platforms aggressively remove misinformation of some kinds, but not all. Iran also has laws restricting which platforms can be used, but does much less to actually enforce these laws. It makes some platforms more inconvenient to use, but does less to actually prevent it. The USA has some new laws restricting the use of social media platforms, specifically Chinese social media platforms. Whether these laws will stand remains to be seen. There are some attempts to control the flow of some types of misinformation on platforms like Twitter and Facebook based on self-regulation by the platform and not government regulation. Throughout our study, we examine the impact of these different approaches on the types of misinformation spreading.

## 2. Background and Related Work

Misinformation is not a new phenomenon, it has been around for centuries in the forms of rumors, gossip conspiracy theories, etc. (Burston et al., 2018). However, the emergence of the publishing industry in the 15th century provided an official modern platform for misinformation. The 21st century has been characterized by the explosion of information through the Internet. The technology, particularly social media, has amplified the spread of misinformation and its adverse impacts by providing a fast and free channel to share any information whether true or false (Lazer et al., 2017).

Since the beginning of the pandemic, catastrophic and life-threatening impacts of tremendous amounts of misinformation circulating on social media have appeared (Donovan, 2020). For instance, due to fake news, some individuals used toxic home remedies resulting in injury and death (Vigdor, 2020a). Misinformation has provoked many people to hoard some vital necessary products (e.g., N95 mask, sanitizers, toilet papers, etc.) causing a shortage of supplies for essential workers (Vigdor, 2020b). Misinterpretations of facts have caused people to not take quarantine seriously and ignore CDC and WHO's recommendations (Centers for Disease Control and Prevention [CDC], 2020a). Within the USA, the Black and African American community has had to deal with misinformation claiming that darker skin may help protect against COVID-19 (Kertscher, 2020) when in fact, risk factors are higher (Farmer, 2020). Other minority groups such as Asian Americans have experienced increased discrimination since the outbreak of COVID-19 (Ruiz et al., 2020). Within India, the Twitter hashtag CoronaJihad exacerbated already present islamophobia as it accompanied false claims of Muslim people intentionally infecting Hindu people (Chaudhuri, 2020; Perrigo, 2020). In some Muslim countries such as Iran and Somalia, religious figures and hardliners believe that true Muslims are immune to the new virus. The religious community is biased against less religious or non-believers and blames them for the pandemic as a form of God punishments (Al Arabiya English & Judd, 2020; Lubrano, 2020; Malekian, 2020). Within China, in the early of this pandemic, one piece of misinformation, widely forwarded through Chinese social media, stated that COVID-19 only attacked Asian people because it was a biological weapon designed to target Chinese people (Steinmetz, 2020; The Storm Media, 2020).

Why is this COVID-19 misinformation so impactful? Desperation and severe stress caused by fear of death, the uncertain nature of the pandemic, economic crisis, unemployment, strict quarantine, and the new routines (e.g., working from home, sanitizing groceries, etc.) has frustrated and distracted billions of people around the world making them even more susceptible to the influence of misinformation(CDC, 2020b; Palsson et al., 2020). Many people are willing to grasp any information (true or false) that might make them feel safe or comfortable or offer easy-to-understand explanations for the complicated pandemic situations (Douglas et al., 2017). On the other hand, some people may boycott receiving information to circumvent such severe stress. Unfortunately, this practice is just as dangerous because being uninformed makes individuals more vulnerable to becoming misinformed (Lazer et al., 2017) and believing false information.

There are a few studies investigating COVID-19 misinformation on social media and more are in development. Kouzy et al. (2020) focused on quantifying the misinformation, but they focused on only English misinformation posted on Twitter (Kouzy et al., 2020). Also, Brennen et al. (2020) proposed some classification of COVID-19 misinformation although their samples are limited to only English language.

The closest related work to this research is the 2020 Misinfodemic Report developed by Meedan and written by Alimardini et al (Alimardani & Elswah, 2020). This qualitative report considers a more global response to misinformation, showcasing seven countries compared to several United States-only studies. Meedan divides their report into the crumbling of public trust, informal leaders of information, and impact of the infodemic on governance. Their coverage of countries is extensive, but our research adds a balance of qualitative and quantitative analysis comparing countries, languages and platforms within our sample.

Purely quantitative research that involves misinformation globally has been done by Pew Research. They surveyed social media users within several countries about how often they encounter obviously fake content (Silver, 2020). This study tries to point researchers toward countries that seem to be encountering misinformation more frequently. This type of quantitative research does not dive as deeply into the content, topics and root of misinformation present within each country surveyed.

## 3. Platforms in Iran, China, and the USA

Twitter, Facebook, Instagram, WhatsApp, Weibo, WeChat and TikTok are the most popular social media platforms in China, Iran, and the USA. However, due to the censorship and political reasons, some strict restrictions have been imposed on these platforms in some cases (Rachman, 2020). Table 1 summarizes the current landscape.

| Platforms | China | Iran | U.S.A. | # of Monthly Active Users (as of 2020) |
|---|---|---|---|---|
| Twitter | ✘ | ✘✓ | ✓ | 330 million |
| Facebook | ✘ | ✘✓ | ✓ | 2.7 billion |
| Instagram | ✘ | ✓ | ✓ | 1 billion |
| WhatsApp | ✘ | ✓ | ✓ | 2 billion |
| Weibo | ✓ | S | S | 550 million |
| WeChat | ✓ | S | ? | 1.2 billion |
| TikTok | ✓ | ✓ | ? | 800 million |

Table 1: Social media Platforms - ✓ for commonly used; ✘ for not allowed/blocked; ✘✓ for not allowed officially but accessible, **S** for seldom used, and **?** for the situation to be determined (CIW Statistics, 2020; Clement, 2020a; Clement, 2020b; Lin, Y., 2020; McGarvey, 2020; Richter, 2020).

Twitter has been blocked in both China and Iran for years (Jen, the Privacy Freak, 2015). The Chinese government has very strict technical and legal measures in prohibiting access to Twitter, though a few government officials have special permission to use it for foreign affairs (Mamiit, 2016). Thus, almost all tweets in Chinese are posted by Chinese speakers outside mainland China and are therefore not a good representation of the social media landscape within China. Restrictions on Twitter in Iran are not as severe as in China despite official filtering of Twitter in Iran.

Facebook has been banned in both China and Iran. However, Instagram is not blocked in Iran at this moment and it is actually very popular, especially among youth, with more than 24 million Iranian active users as of January 2018 (Financial Tribune, 2018). Instagram has been banned in China and much like with Twitter and Facebook, this ban is more aggressively enforced.

Weibo (微博) and WeChat (微信) are the most popular social network platforms in China (DeGennaro, 2019). They are not only popular inside China but are also used heavily among Chinese speakers outside of China.
WeChat is beyond a social networking and messaging platform and its monthly active users have reached about 1.2 billion in the second quarter of 2020 (CIW Team, 2020). In addition to social networking services, WeChat offers one of the most popular payment methods in China, called "WeChat Pay" (微信支付). WeChat could be considered as the combination of WhatsApp, Facebook, and PayPal.

WhatsApp is a cross-platform encrypted messaging application acquired by Facebook, and its monthly active users have reached 2 billion as of March 2020 (Clement, 2020b). It has 68.1 million users just in the USA as of 2020 (Andjelic, 2020). WhatsApp is completely banned in China. WhatsApp is legally allowed to be used in Iran and has become even more popular in Iran after the Iranian government blocked Telegram in 2018 which used to be the most popular messaging application in Iran. The government claimed that Telegram had endangered national security (Erdbrink, 2018).

WhatsApp has become a makeshift social media platform in Iran as group chats have begun including thousands of members forwarding and sharing information. Although the government has blocked many social media platforms officially, many Iranians still use VPN and Proxy anti-filter apps/tools to access blocked social media. However, this access is more limited and requires some technical tools and skills. This has helped to drive usage towards WhatsApp which is a private end-to-end encrypted messaging platform without tracing capabilities. Partly as a result, WhatsApp has become a major source of misinformation in Iran and many other countries (Alimardani & Elswah, 2020). WhatsApp recently limited the number of times users can forward a message to only five times, in an attempt to fight against misinformation (Kastrenakes, 2019).

TikTok was developed by ByteDance (字节跳动), a Chinese company in Beijing. TikTok has a version used only in China called Douyin (抖音) to separate the domestic users from international users. Due to the concern of cyber security, TikTok operation in the USA will be transferred to a new company named TikTok Global and will cooperate with Oracle and Walmart to ensure the data safety (Lin, L., 2020). TikTok is allowed to be used in Iran.

### 3.1. Search Methods in Different Languages

Keywords and hashtags are the major methods to search on social media and we use both to collect our data set. All platforms named in this study supports both keywords and hashtag search. Hashtags on WhatsApp might not be as popular as other platforms however this function makes the search process very easy even through a private message chain (Patkar, 2013).

Unlike hashtags on American platforms, a hashtag on Weibo is owned by a host. Also, each hashtag has its own unique webpage, which is called 超级话题 (i.e., topic or super topic). Since January 2020, many super topics have appeared around COVID-19 on Weibo. Figure 1 shows a sample of super topic webpage on Weibo. The topic host and the largest contributor to this topic is a state media. Although English hashtags are allowed on Weibo, almost all hashtags are in simplified Chinese. Weibo needs the symbol "#" before and after a term to function as hashtags, unlike Twitter which only needs "#" before. WeChat Channels is a popular feature of WeChat and contain public feeds of content. Hashtag and keywords search can be used in WeChat Channels.

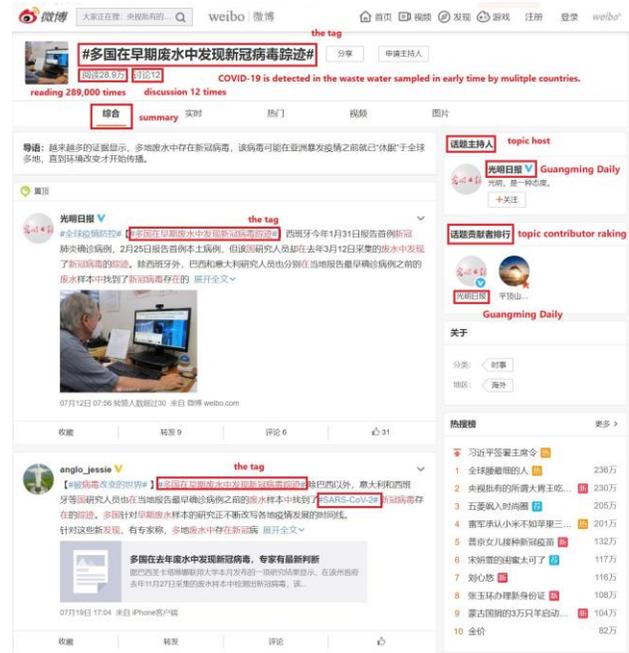

Figure 1: A Chinese Weibo super topic #多国在早期废水中发现新冠病毒踪迹#, translation: "COVID-19 is detected in the wastewater sampled in early times by multiple countries".

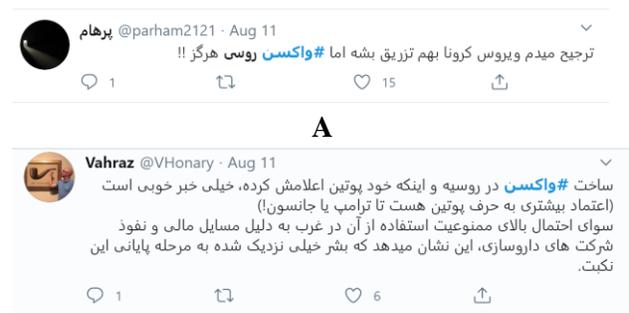

Figure 2: A Farsi hashtag #واکسن (Vaccine) on Twitter. A) translation: "I prefer to get Coronavirus rather than get a Russian Vaccine". B) translation: "Russian vaccine announced officially by Putin is a great news. Putin's promises are more reliable than Trump's or Boris Johnson's."

Iranians use both English and Farsi hashtags across different platforms. An interesting observation is that the same hashtag can be used for two completely opposite opinions. For instance, #واکسن is a hashtag about the Russian vaccine used to criticize the vaccine (Figure 2-A) and to express excitement about it (Figure 2-B).

## 4. Our Methodology

Selecting an appropriate sampling methodology is a major challenge when it comes to social media studies. Non-biased sampling from social media is often difficult because the dataset is so highly dynamic, massive in size, and difficult to filter. The sheer volume of data--Facebook has over 350 million users and Twitter has a rate of 17,000 tweets per minute--makes the gold standard of data acquisition, true random sampling, challenging.

This can be even more complicated when sampling for misinformation since our underlying population is not all posts, but those posts that contain misinformation that has been refuted or debunked by recognized fact-checking organizations. Hashtag and keyword filters do not necessarily guarantee to find misinformation, and even if a suspected post is found, we are reliant on the efficiency of journalists in fact-checking posts and content which is also biased by the particular needs and intentions of journalists and respective media outlets (Pingree et al., 2018).

In this study, we utilized an opportunistic sampling strategy, meaning that our sampling was determined by the population, or presence of misinformation on social media, that was available and officially debunked at the time, and our abilities to find them. We used Chinese, English, and Farsi hashtag and keyword filters to collect 200 specific items of debunked misinformation that spread virally between January 1 and August 31. When we encountered a particular item in one of our languages, first we used fact checking organizations to see if that claim had been debunked using fact-checking sources such as International Fact Checking Network (IFCN), platform based fact-checking tools in Twitter, Facebook and Google, and organizations using Claim Review (Poynter, n.d.; Schema, 2014). The item is logged only if we could find a verified debunking source; otherwise, it is discarded. We estimate that number of logs could be doubled if we did not have the debunking constraint. Then we watched for posts with the same claim in our other languages and on other platforms. We also sought out independent media reports beyond our own direct experience that the referenced claim was spreading virally. Thus, each of these 200 items of misinformation in our study represents thousands of posts repeating the same debunked claim, often across multiple languages and platforms.

Out of these 200 collected pieces of misinformation, 54 are in Chinese, 156 in English, 111 in Farsi. We admit that our data set may be unintentionally biased considering the biased nature of social media and the fact that all authors are currently in the USA, so we might be more exposed to English misinformation. Furthermore, due to the lack of copyright laws enforcements in Iran, a single piece of misinformation might have been repeatedly debunked by too many sources which prolongs and complicates the search process for new pieces. Figure 3 represents a clearer breakdown of our collected data and the overlaps in misinformation across languages.

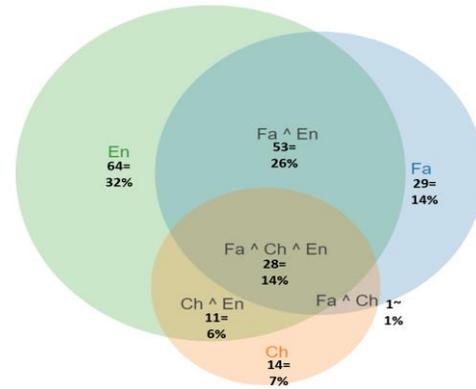

Figure 3: Overlaps of collected data across the multiple languages. Ch, En, and Fa refers to Chinese, English, and Farsi. Total are 200= 29(Fa)+ 14(Ch) +64(En) +53(En^Fa)+11(En^Ch)+1(Fa^Ch)+28(Fa^Ch^En).

Strict filtering and censorship policies for Chinese social media is a major factor in the notably lower totals. This can be considered as a benefit of the aggressive censoring of some types of misinformation in China. However, this could come at the expense of silencing voices that are needed to counter other kinds of misinformation. This is a key question being explored in this paper.

There are 14, 64, 29 pieces of misinformation found exclusively in Chinese, English, and Farsi only, respectively. Out of overall 200 pieces of misinformation, 53 occurred in only Farsi and English (26%) which is the largest overlap among possible pairs of languages while Farsi and Chinese have the least overlap (less than 1%). Of the 200 collected pieces of misinformation, 14% have been found in all three languages.

## 5. Topics of COVID -19 Misinformation

Given the complicated nature of tracking misinformation across multiple languages' social media landscapes, a comprehensive categorization over the topics of COVID-19 misinformation was critical to analyze the catastrophic infodemic occurring during this global pandemic. We identified 10 top level categories for the topics of COVID-19 misinformation. We found that these proposed categories were inclusive enough to cover the Chinese, English, and Farsi misinformation in our collection. These categories are also extendable to also cover the possible future misinformation. The description of these categories are as follows:

1. **Cures:** includes traditional, superstitious, fake, or ineffective treatment methods, products, remedies and claims, etc.
2. **Origin:** includes claims about how the virus originated in the first place.
3. **Testing:** includes topics such as availability of testing, medical testing kits created by different countries, unconventional and non-medical methods for testing, testing cost, etc.
4. **Vaccines:** includes topics such as safety and effectiveness of vaccines, competition over the first vaccine, the length of the immunity that vaccine can generate, mass production, implementation of vaccination, etc.
5. **Prevention methods (public):** includes topics concerning public and macro policies and strategies to prevent virus transmission.
6. **Prevention methods (individual):** includes traditional, superstitious, fake, or ineffective methods of prevention identified as individual or personal actions.
7. **Number of deaths and confirmed cases (Statistics):** includes rumors, actions, and false claims to manipulate the official statistics of death and confirmed cases including exaggerating or downplaying the numbers.
8. **Rumors about other countries (often xenophobic rumors internal and external to a country):** includes conspiracy theories and rumors spread out about other countries' roles related to the new virus.
9. **Virus transmission:** includes topics related to misleading, superstitious, or fake, methods by which the virus can transmit, asymptomatic period, basic reproduction number (R0).
10. **Others:** includes topics such as contact tracing (pros and cons of contact tracing, and rumors about the amount of personal information needed to be collected for the contact tracing purposes); recovery (the length of recovery periods, antibody level after the recovery, immunity of recovered patients, etc.); prediction of the pandemic; compensations, and more topics; etc.

The topics for each of our collected samples were inductively examined to be specific to COVID-19 misinformation and accordingly classified within the proposed 10 categories. Figure 4 represents the proportion of each topic within our overall collection across languages.

About a quarter of the total collected misinformation in our sample fell under the topic of prevention-individual. Also, the top three topics of the collected misinformation mostly concern the individual behavior. In a few cases, the same claim has been reported in more than one category, e.g., "drinking bleach" has been circulating on social media as both a cure and prevention method. In fact, this high proportion of misinformation within the context of individual behavior verifies the vulnerability of users to misinformation to cope with uncertainty and uncontrollability of pandemic circumstances.

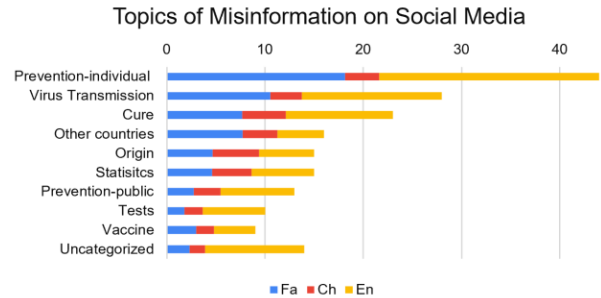

Figure 4: Proportion of sample (n=200) within the proposed topics categories.

An interesting observation on the topics of misinformation in these three languages is that the percentage of categories in English and Farsi are comparable. Furthermore, the top two categories for both Farsi and English are "Prevention- Individual" and "Virus Transmission". This suggests that the Iranian laws coupled with weak enforcement still allows misinformation of a similar kind to flow in Iran as in the USA with fewer laws and restrictions.

The greatest difference in English and Farsi belongs to the category of misinformation related to "Other Countries". Meanwhile, Chinese and Farsi misinformation within this category have similar records. Since most of the Chinese and Farsi rumors about other countries have political roots and considering the controlling governments in Iran and China, such similarity is not surprising. It is interesting that despite tight control of other kinds of misinformation in China, this category is still specifically allowed to flow. The top category of the Chinese sample belongs to the topic of "Origin" of the virus which is actually still a question mark for the world.

We also observed a larger variety of topics in English which explains the relatively higher percentage of English misinformation categorized as "Others". For instance, misinformation related to contact tracing could only be found in the USA. One of the reasons could be the fact that some laws and rules vary from one state to another in the USA (*e.g.*, misinformation about federal and state compensation and financial support from businesses and individuals) and this confusing landscape of varying laws opens up space for misinformation to flow.

## 6. The Roots of Misinformation

In this section, we delve beyond a classification of misinformation by topic to the roots of misinformation. For example, Jang et al. reported that most misinformation stems from a false statement quoted by a public figure; or is deliberate misinformation used for a particular purpose (Jang et al, 2018). We found that COVID-19 misinformation also followed this same pattern.

We identified the following six categories for the roots of the misinformation. It should be noted that we use the term "root", but we could have instead used terms such as "reason" for or "source" of misinformation.

1. **Political-related roots:** a false statement quoted by a political figure; or related to governments and the relationship between countries; or used for political purposes, e.g., elections.
2. **Medical/Science-related roots:** a false statement quoted by someone claiming to be a medical expert, e.g., doctors, nurses, etc.; or a false perception related to medical research outcomes.
3. **Celebrities & Pop Culture-related roots:** a false statement quoted by a celebrity, influencer, or popular/public figure in the media field; or, a misleading/false content such as a video, photo or an article gets viral through media, e.g., TV, press, etc.
4. **Religious-related roots:** a false statement quoted by a religious figure; related to religious and/or traditional and superstitious beliefs.
5. **Criminal-related roots:** a false statement which has been claimed by a scammer or hackers for criminal purposes such as fraud, access to personal information.
6. **Others:** any other false statements that cannot be substantiated to be related to the mentioned categories including hoaxes, jokes and other undesignated misinformation.

For each of our 200 pieces of misinformation, we attempted to track the source that led to this content going viral. The roots of misinformation collected in our sample are verified by reliable references and categorized in one of these six categories by human annotators. If a false statement has been quoted by multiple sources, then the source which made the statement goes viral will be considered as the root of the statement. As Figure 5 represents, more than one third (33.5%) of the misinformation has been related to political roots which is alarming and shows the critical role of governments and political figures in the infomedic. For example, researchers analyzing 38-million English-language articles about the pandemic found that USA President Trump was the largest driver of the infodemic over that sample (Evanega et al., 2020). We could not verify the roots of 24% of our sample, indicating the breadth and diversity of COVID-19 misinformation; these have been categorized as "Others".

The three countries selected in this paper have their own unique characteristics and cultural structures and here we further discuss the roots of misinformation, particularly the religious and political roots, separately in Iran, China, and the USA in the following subsections as well as a discussion of commonalities across these countries.

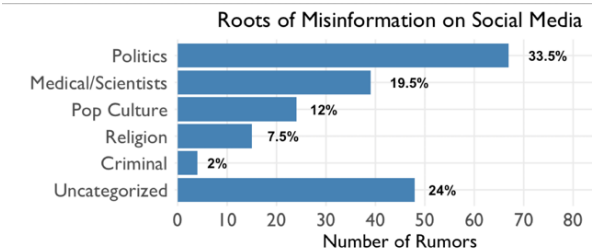

Figure 5: Proportion of the collected data (n=200) across the proposed categories for roots of COVID-19 Misinformation.

### 6.1. Roots in China

The COVID-19 misinformation circulating on Chinese social media landscape reflects significant differences in political systems between China and the western world. For such reasons, misinformation from the western world is translated, filtered, and reflected in Chinese social media. A primary root of misinformation circulation in China is the presence of fake science-based claims.

When considering Chinese social media, it is important to consider the systems of aggressive censorship in China that results in a strict politically filtered internet. Global misinformation which has been translated and reflected in Chinese social media environment can be considered politically biased since the vast majority of social media in China is tightly controlled by the state.

An interesting trend observed in Chinese social media is that some already debunked misinformation from the rest of the world, especially the USA, has been translated and widely shared in Chinese platforms to criticize the western world's attempts to fight COVID-19. This pattern of disclosing the debunked misinformation is a politically manipulated misinformation per se. Thus, we can only call this type of message pseudo "misinformation" (e.g. reports of President Trump suggested injecting bleach to treat COVID-19). In this case, it appears that the reason this misinformation is allowed to flow is more of a criticism of the USA and the susceptibility of its citizens to misinformation that appears clearly ridiculous to the average Chinese citizen. Through translating and sharing this message, people in China have been shocked to see such clear misinformation from the USA President and will further believe in more similar information regarding the unsuccessful control of COVID-19 in the USA.

Much of the true misinformation in China are fake-science misinformation based on some assumption without sufficient evidence and scientific work to support it. Though a post may point to scientific sources to back up the claim, it is difficult for most people to verify the source due to the language barrier. Specifically, most people do not have the capacity of reading research publications in English or even of translating the name of the journal. This style of post can be used to give the impression of a scientific evidence that may not exist. In our Chinese samples, the largest category of roots of misinformation belongs to politics (41%) which is aligned with our analysis for English and Farsi samples (Figure 6-A).

Due to tight controls and censoring in China, collecting samples of COVID-19 misinformation in Chinese has been relatively harder than English and Farsi. Also, identifying the roots of the collected COVID-19 misinformation is complicated. The roots of 24% of the Chinese misinformation remain undesignated and categorized as "Others". Surprisingly, a large proportion of COVID-19 misinformation in Chinese has been started by celebrities and/or pop culture (20%) which is larger than the English and Farsi corresponding category.

Unlike the Iran and USA, most Chinese are irreligious and atheist (Huff Post & Briggs, 2011; Noack, 2015). However traditional Chinese beliefs have been identified among COVID-19 misinformation in Chinese (only 2%) and categorized as having religious roots.

We could not identify any misinformation in Chinese associated with the criminal roots (0%). This is a strong result and could be considered one benefit of tight government control over social media in China. Confronted with the impact of misinformation on western democracies (e.g. advice to drink bleach), Chinese citizens could be convinced that the aggressive censorship policies help keep their society safer for criminal elements and viral misinformation.

### 6.2. Roots in the USA

Misinformation in the USA is commonly in the form of counter-expertise, the rejection of mainstream academic expertise, which dates back to the 19th century (Douglas, 2018). This form of misinformation began with Christian fundamentalists rejecting evolution as it contradicted the Bible (Douglas, 2018). Since then, counter-expertise looks to distrust mainstream scientific media by promoting alternative thoughts through alternative media. This alternative media happens to be far more susceptible to misinformation (Douglas, 2018). Such alternative media has released misinformation throughout the pandemic, with Fox News ("Fox's Dr. Marc Siegel", 2020) stating "the virus should be compared to the flu because at worst case scenario it could be the flu".

Across the USA, churches have successfully resisted complying with government-led preventive measures and health orders. A recent study shows that 71% of Protestant/Evangelical ministers held in-person worship as of July 15, 2020 (Vondracek, 2020). The persistence to stay open and hold in-person worship has led to churches becoming hot spots for positive cases (Conger et al., 2020). Father Joseph Illo, leader of the Star of the Sea Church, sent out misinformation to churchgoers stating that "the news reports about COVID are largely unreal" (Bisacky, 2020). However, not all churches within the USA are rejecting mainstream media or health orders.

The current politically polarized atmosphere in the USA is also a major driver of misinformation. Misinformation related to politics is often fabricated to create confirmation bias among readers, subsequently leading to heightened in-group/party identification and polarization (Douglas, 2018) (e.g. "The political party I identify with fought for the right thing, as opposed to the other party").

COVID-19 misinformation spread by political figures in the USA may be a symptom of deflection and scapegoating due to inadequate administrative response. USA government sources have suggested a conspiracy theory that COVID-19 was created in a Chinese laboratory (Aljazeera news, 2020). Deflection is apparent as the USA President Donald Trump suggested "quick fixes" and "cures", such as hydroxychloroquine, azithromycin, and convalescent plasma (Caplan, 2020), that have limited testing and results as potential cures or "game changers in the history of medicine".

It is necessary to acknowledge the severity and seriousness of the spread of misinformation from the USA public officials. In our collection of misinformation samples from the USA about one third (31%) have political roots (Figure 6-B). While the spread of information from public officials to win elections is not new, the role of the mass media as a corrective measure on this behavior has changed. The media is now competing with social media for advertisements, which has resulted in pressure to cater content to users, further perpetuating political polarization (Scheufele & Krause, 2019). From drastically downplaying the seriousness of the virus to stating that "one day it's like a miracle, it will disappear" (The White House, 2020), the USA Government has contributed heavily to COVID-19 misinformation on all media platforms.

The role of celebrities in driving American culture (Clemmons, 2018) and the power of American public figures' words should not be underestimated. In our sample, 13% of the collected misinformation in the USA has been directly traced back to a celebrity, influencers or popular figure. For example, on July 31, 2020 Instagram removed Madonna's post "for making false claims about cures and prevention methods for COVID-19" (Solis, 2020). False statements by self-claimed medical related crew or wrong and manipulated interpretation from medical facts are another major root of misinformation in the USA (22% of our sample represented in Figure 7-B).

### 6.3. Roots in Iran

COVID-19 misinformation has hit Iran harshly. Two major reasons have been identified for creating and spreading misinformation: first, discourse about COVID-19 is politically manipulated by the government (Alimardani & Elswah, 2020); second, official religious figures have interfered with COVID-19 related issues and religion has been a barrier for ordinary hardliners to be unbiased. In general, social media reflects that people do not trust the COVID information released by either government or religious officials. When the official channels of communication of information fail, people start clinging to their own unofficial channels of information gathering without monitoring the validity of the information which eventually intensifies the spread of misinformation.

Iran has a controlling and conservative government which micro-manages all aspects of people' lives. This characteristic of the government politicizes every subject including the pandemic. Thus, it is not surprising that much of the COVID-19 misinformation found in Farsi has political roots. An ironic piece of misinformation with political roots in Iran is that the government believes the COVID-19 misinformation present in Iran has been started mainly by "enemies", referring to the USA government ("BBC News Persian", 2020).

Another example of political misinformation is that the Iran government promoted a fake testing technology called "coronavirus remote detectors" which can detect infected individuals from a distance of 109 yards. The unveiling ceremony on April 15th went viral all-over the Iranian press and social media.

Religion has also played a critical role in spreading COVID-19 misinformation in Iran. For instance, some official religious hardliners falsely believe that sacred protection from religious shrines would prevent infection (Malekian, 2020). In Iran, some Shiite Muslim religious figures often use people's faith to oppose "westernized" facts, sometimes including scientific facts. For example, on February 24, 2020, a religious figure, Ayatollah Abbas Tabrizian, advised people to rub their anuses with violet oil to prevent and cure COVID-19. This post on his official Telegram channel (with more than 200,000 followers) has been viral on all Farsi social media. Users reshared this post with mixed reactions that included both adherence and ridicule (The New Arab, 2020).

As it was expected, in our collected Farsi sample, the top category of roots of misinformation belongs to politics (27%) (Figure 6-C). The next largest proportion of the misinformation has medical/science roots including both western misinformation and local and traditional Persian remedies. About 11% of our sampled misinformation has religious roots. Considering the bold role of religion in Iran, this rate seems relatively low. However, the virality of this misinformation has been substantial, such that some of the misinformation is still circulating on social media, even after it officially got debunked. Some however have been transformed into sarcasm to be used as a form of protest against hard-core religious figures.

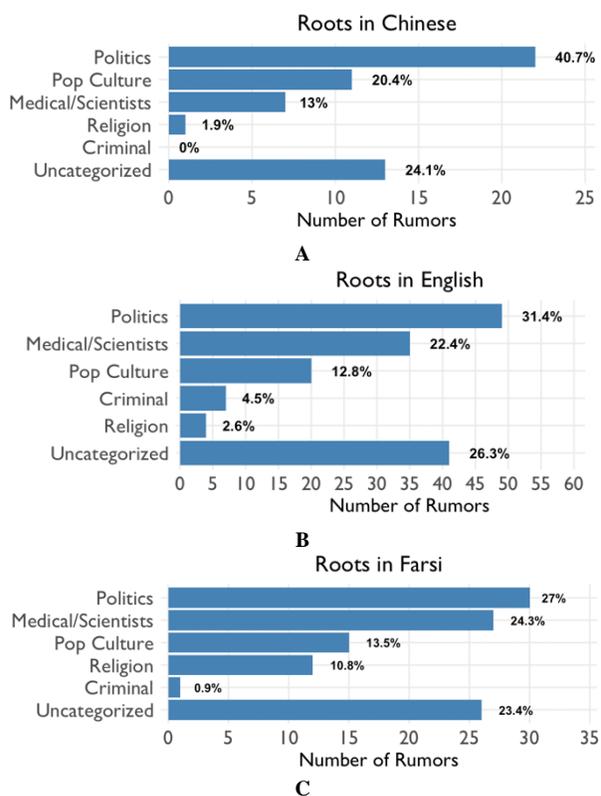

Figure 6: Proportion of the Chinese (n=54), English (n=156), and Farsi (n=111) collected data across the proposed categories for roots of COVID-19 Misinformation.

### 6.4. Discussion

A recent study showed that there is an intersection between fake news and religion in societies with religious background (Douglas, 2018). Our observation on this matter in Iran is aligned with this study given the Islamic background of Iran (11%). However, our sample could not confirm the same result in the USA and China. Given the impact of the Christian community in the USA (Bailey, 2019), 3% religious roots for the English misinformation was surprisingly lower than our initial expectation.

When politics and government play a significant role in the destiny of a society (which is the case in Iran, China, and the USA), a political polarization phenomenon will emerge (Facing History and Ourselves, 2020). Political polarization has been known as an important factor to spread misinformation in a society and the meaningful relationship between misinformation and political polarization has been profoundly investigated in the literature (Donovan, 2020; Pennycook & Rand, 2019). Polarization encourages rigid opinions and bias against opposite groups (Interian & Ribeiro, 2018); Lelkes, 2016). Consequently, it can cause misinterpretations of facts and the spread of fake news, misinformation, conspiracy theories, etc. (Pennycook & Rand, 2019). Our results in Sections 6 also support that politics and political polarization are the roots of the majority of Chinese (41%), English (31%), and Farsi (27%) collected samples of COVID-19 misinformation.

The absence of misinformation with criminal roots in Chinese social media is notable. This is an example of a key difference in how government strategy directly influences the types of misinformation to which the public is exposed. As a result, some societies are more vulnerable to criminal misinformation. Given that politics was the largest root of misinformation across all three languages in this dataset, all societies are extremely vulnerable to government misinformation. However, some have more potential to counter government misinformation with information from private sources. In many ways, this is perhaps the key question for countries and societies around the world going forward in deciding how they want to control misinformation and infodemic. For liberal democracies, a key challenge is determining how to control misinformation without silencing the voices needed to hold government misinformation accountable.

## 7. Conclusions

Our major goal is to analyze the COVID-19 misinformation on different social media platforms across different languages to gain a more holistic, global understanding of the misinformation's landscape. This effort is an initial step to diminish the current infodemic happening along with the pandemic. By increasing public knowledge of the adverse impacts of misinformation on public health during the pandemic, many lives could be saved.

To achieve our goal, the opportunistic sampling approach was utilized to compile 200 pieces of verified misinformation posted virally in Chinese, English, and Farsi across Twitter, Facebook, Weibo, WeChat, WhatsApp, Instagram, and TikTok between January 1 and August 31. Each of these 200 pieces represented thousands of posts across platforms and often across languages. Then, a classification approach was proposed to categorize the collected misinformation based on both their topics and roots. We identified 10 high level topics being inclusive and relevant in all three languages. We also identified 6 major categories for the roots of misinformation. Our study yielded the following important results:

- Politics was the largest root of misinformation across all three languages in this dataset.
- Overall, the English and Farsi samples have more in common in terms of the topic of misinformation than Chinese specifically regarding individual prevention methods.
- The absence of misinformation with criminal roots and fewer categories of misinformation overall in Chinese social media is notable and points out a critically important tradeoff in the control of misinformation.

We note important differences in how government controls on social media platforms drive usage onto some platforms and away from others, with different infrastructure for tracking and controlling misinformation. Understanding how different countries utilize social media and their restrictions gives better insight as to how to regulate disruptive behavior. A key challenge going forward for all societies and countries will be in determining how to control misinformation without silencing the voices needed to hold governments accountable. Overall, it is clear how focusing beyond English, beyond the USA, and beyond the USA-based social media platforms are essential to providing a clear understanding of the effects of misinformation and the effectiveness of misinformation control strategies around the world.

## 8. Acknowledgment

The authors would like to kindly acknowledge assistance from Gillian Kurtic and Yan Chen for contributions to the paper including help reviewing, organizing, and formatting. We also thank Dr. Ricardo Baeza-Yates for his detailed feedback, wise advice and thought-provoking questions. We gratefully acknowledge funding from Clarkson University Epidemic and Virus-Related Research Innovation fund.